\documentclass[fleqn,twoside]{article}
\usepackage{espcrc2}
\usepackage{graphics}
\def\lsim{\raise0.3ex\hbox{$<$\kern-0.75em\raise-1.1ex\hbox{$\sim$}}}
\def\gsim{\raise0.3ex\hbox{$>$\kern-0.75em\raise-1.1ex\hbox{$\sim$}}}

\newcommand{\pslash}{p\kern-1ex /}
\newcommand{\Dslash}{{\cal D}\kern-1.5ex /}

\newcommand{\beqa}{\begin{eqnarray}}
\newcommand{\eeqa}{\end{eqnarray}}

\newcommand{\beq}{\begin{equation}}
\newcommand{\eeq}{\end{equation}}
\newcommand{\bc}{\begin{center}}
\newcommand{\ec}{\end{center}}

\title{
\vspace*{-2.cm}
\begin{flushright}
{\normalsize UTHEP-475}\\
{\normalsize UTCCP-P-141}\\
\end{flushright}
$I=2$ pion-pion scattering phase shift in the continuum limit 
calculated with two-flavor full QCD 
\thanks{Talk presented by T.~Yamazaki}}
\author{CP-PACS Collaboration :
  T.~Yamazaki\rlap,\address{Institute of Physics,
    University of Tsukuba, Tsukuba, Ibaraki 305-8571, Japan}
  S.~Aoki\rlap,$^{\rm a}$
  M.~Fukugita\rlap,\address{Institute for Cosmic Ray Research,
    University of Tokyo, Kashiwa, Chiba, 277-8582, Japan}
  K-I.~Ishikawa\rlap,\address{Department of Physics,
    Hiroshima University, Higashi-Hiroshima, Hiroshima
    739-8526, Japan}
  N.~Ishizuka\rlap,$^{\rm a,}$\address{Center for Computational Physics,
    University of Tsukuba, Tsukuba, Ibaraki 305-8577, Japan}
  Y.~Iwasaki\rlap,$^{\rm a,d}$
  K.~Kanaya\rlap,$^{\rm a}$
  T.~Kaneko\rlap,\address{High Energy Accelerator Research Organization
    (KEK), Tsukuba, Ibaraki 305-0801, Japan}
  Y.~Kuramashi\rlap,$^{\rm e}$
  V.~Lesk\rlap,$^{\rm d}$
  M.~Okawa\rlap,$^{\rm c}$
  Y.~Taniguchi\rlap,$^{\rm a}$
  A.~Ukawa$^{\rm a,d}$ and
  T.~Yoshi\'e$^{\rm a,d}$
}

\begin{document}
\pagestyle{empty}

\begin{abstract}
We present a calculation of 
the scattering phase shift for the $I=2$ S-wave 
pion-pion system in the continuum limit with two-flavor full QCD.
Calculations are made at three lattice spacings, using  
the finite volume method of L\"uscher in the center of mass frame, and 
its extension to the laboratory frame.
\end{abstract}

\maketitle

\section{Introduction}
\label{sec:intro}

Pilot studies of the scattering phase shift were carried out
for the $I$=2 S-wave two-pion system in Ref.~\cite{Fiebig,CP-PACS:phsh}.
These studies were made only at one lattice spacing 
employing the quenched approximation.
In this report, we present a calculation including  
two flavors of dynamical quarks, so that we can
get rid of errors from unitarity violation,
at three different lattice spacings
to carry out extrapolation to the continuum limit.

Our study is based on 
the finite volume method of L\"uscher formulated for the center of mass
(CM) frame~\cite{FVM:L}, and its extension to the laboratory 
frame by Rummukainen and Gottlieb~\cite{FVM:RG}. While the inclusion of
the lab frame adds little extra computational cost,  
the use of the lab frame allows significantly more dense sampling  
of the energy states.

We use the full QCD configurations~\cite{CP-PACS:sprm} 
previously generated with 
an RG-improved gauge action and a mean-field improved clover quark action
at three gauge couplings $\beta = 1.80, 1.95$
and $2.10$, corresponding to $a^{-1} \approx 0.92, 1.3$ and
$1.8$ GeV, on 
$12^3 \times 24, 16^3 \times 32$
and $24^3 \times 48$ lattices.
The physical volumes are about $2.5$ fm.
Four hopping parameters corresponding to
$m_{\pi} \approx 1.0, 0.9, 0.75$ and $0.55$ GeV
are taken at each $\beta$ for chiral extrapolation.
The numbers of the configurations vary from 380 to 720.

\section{Methods}

On a finite volume of $L^3$ the interaction between two pions
shifts the two-pion energy $E$.
In the CM frame, writing 
$E = 2 \sqrt{ m_{ \pi }^2 + p^2 }$ with 
$p^2 = ( 2 \pi / L )^2 \overline{n}$ ($\overline{n}$ being non-integer),
the phase shift $\delta( p )$ is given by the relation
$
\tan \left( \delta( p ) \right) = 
{ \pi^{ 3 / 2 } \sqrt{ \overline{n} } }/
     { Z_{ 00 } ( 1 ; \overline{n} ) }
$
where $Z_{ 00 }(i;\overline{n}) = 
(1/\sqrt{4\pi})\sum_{\vec{l}}(l^2-\overline{n})^{-i}$,
$\vec{l}$ being integer vectors.

In the laboratory frame, the two-pion energy $E_L$
and the sum of the two pion momenta $P=p_1+p_2$ are 
given by 
$p^2 = ( 2 \pi / L )^2 \overline{m} 
= ( E_L^2 - P^2 ) / 4 - m_{\pi}^2$.
The phase shift is now  
$\tan\left( \delta( p ) \right) = 
{ \gamma \pi^{ 3 / 2 } \sqrt{ \overline{m} } }/
     { Z^P_{ 00 } ( 1 ; \overline{m} ) }$
where $\gamma = E_L/\sqrt{ E_L^2 - P^2 }$ and
$ Z^P_{ 00 } ( i ; \overline{m} ) = 
(1/\sqrt{4\pi})\sum_{\vec{r}}(r^2-\overline{m})^{-i}$.
Here the summation is taken over the vectors $\vec{r}$ that
are constructed as $ \vec{r} = \gamma^{-1}( \vec{l} + \vec{d} / 2 )$
where $ \vec{d} = L \vec{P} / 2 \pi$.
\begin{figure}[t!]
\centerline{\scalebox{.44}[.44]{\includegraphics{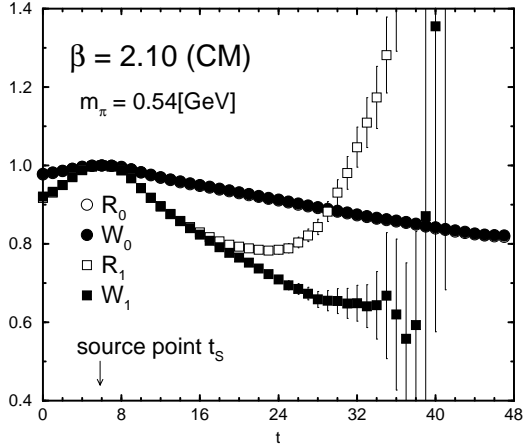}}}
\vspace{-1.3cm}
\caption{Ratio of the four-point to the square of two-point functions.
The values before diaginalization $R_n (t)$ are compared to those
after diaginalization $W_n(t)$. Note that $W_0(t)$ is completely
overlayed on $R_0(t)$.
\label{fig:diag}}
\vspace{-8mm}
\end{figure}

The pion four-point function behaves 
as a sum of exponentials 
due to the presence of a number of 
states having the same quantum numbers. 
To resolve the energy eigenvalues, 
we construct in each frame the pion four-point function matrix 
$C_{nm}(t) = \langle 0 | \Omega_n(t)
                         \Omega_m(t_S) | 0 \rangle $,
where $t_S$ is the source point and 
$\Omega_n(t) = \pi(p_1)\pi(p_2)$ is the two-pion operator
with momenta $p_i$ that depends on the state $n$. 
The energy eigenvalues $\lambda_n(t) = {\mathrm {exp}}[ - E_n(t-t_0)]$
are obtained by diagonalizing the matrix 
$C^{-1/2}(t_0) C(t) C^{-1/2}(t_0)$ with
$t_0$ a reference time~\cite{DIAG}.
A cut-off  $N$ is introduced in the 
number of the energy states considered. 

We work with the CM and two laboratory
frames, L1 and L2.
The sums of the two pion momenta $\vec{P}$ in L1 and L2 are 
$\vec{P}_{\mathrm L1}=(1,0,0) \times (2\pi/L)$ and
$\vec{P}_{\mathrm L2}=(1,1,0) \times (2\pi/L)$.
We choose $N=3$ in the CM frame and $4$ in the lab frame.
In each frame, $3\times 3$ or $4 \times 4$ matrices are diagonalized
and the phase shift is determined for 
the energies of the ground ($n=0$) and the first excited ($n=1$) states.

\begin{figure}[t!]
\centerline{\scalebox{.40}[.40]{\includegraphics{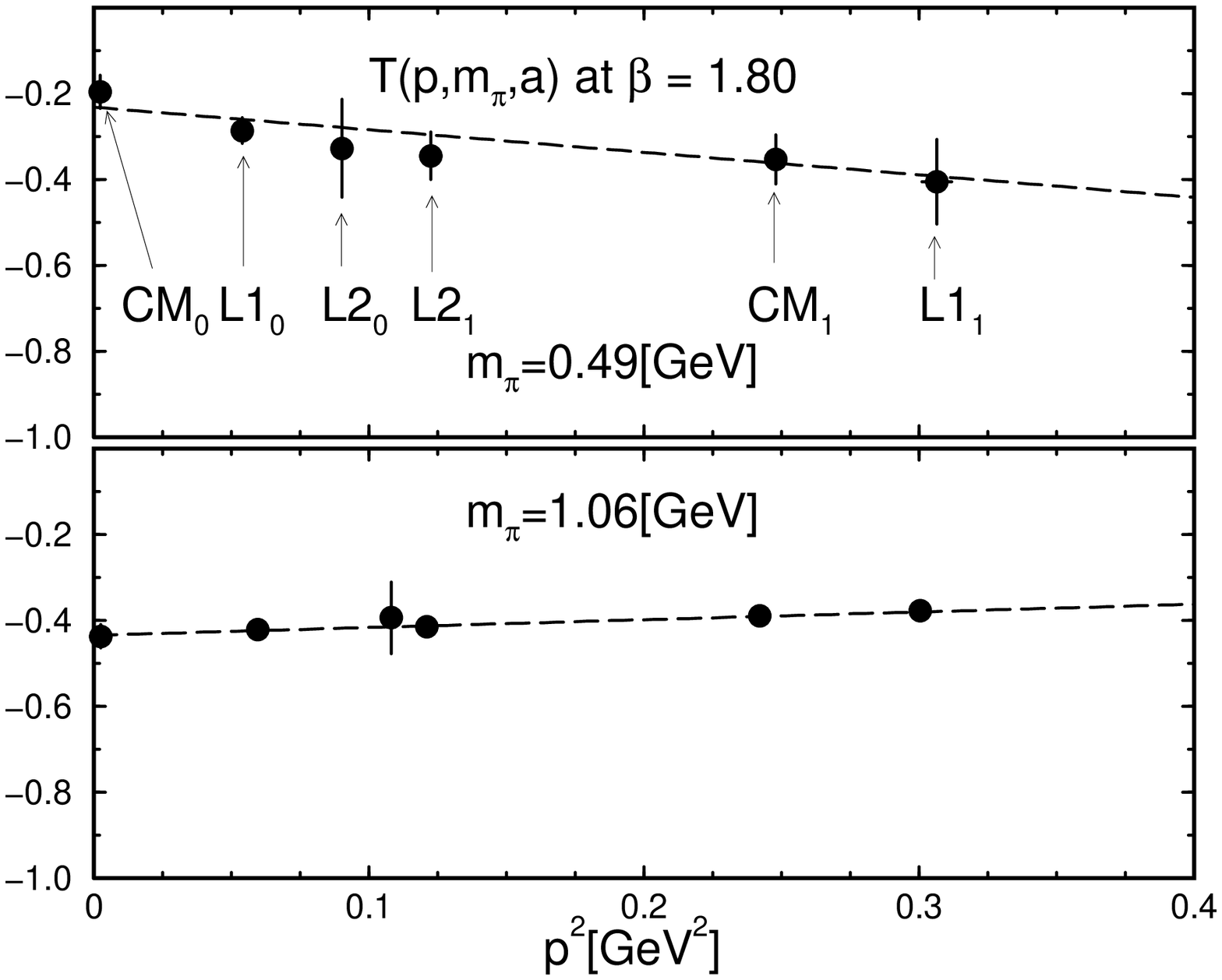}}}
\centerline{\scalebox{.40}[.40]{\includegraphics{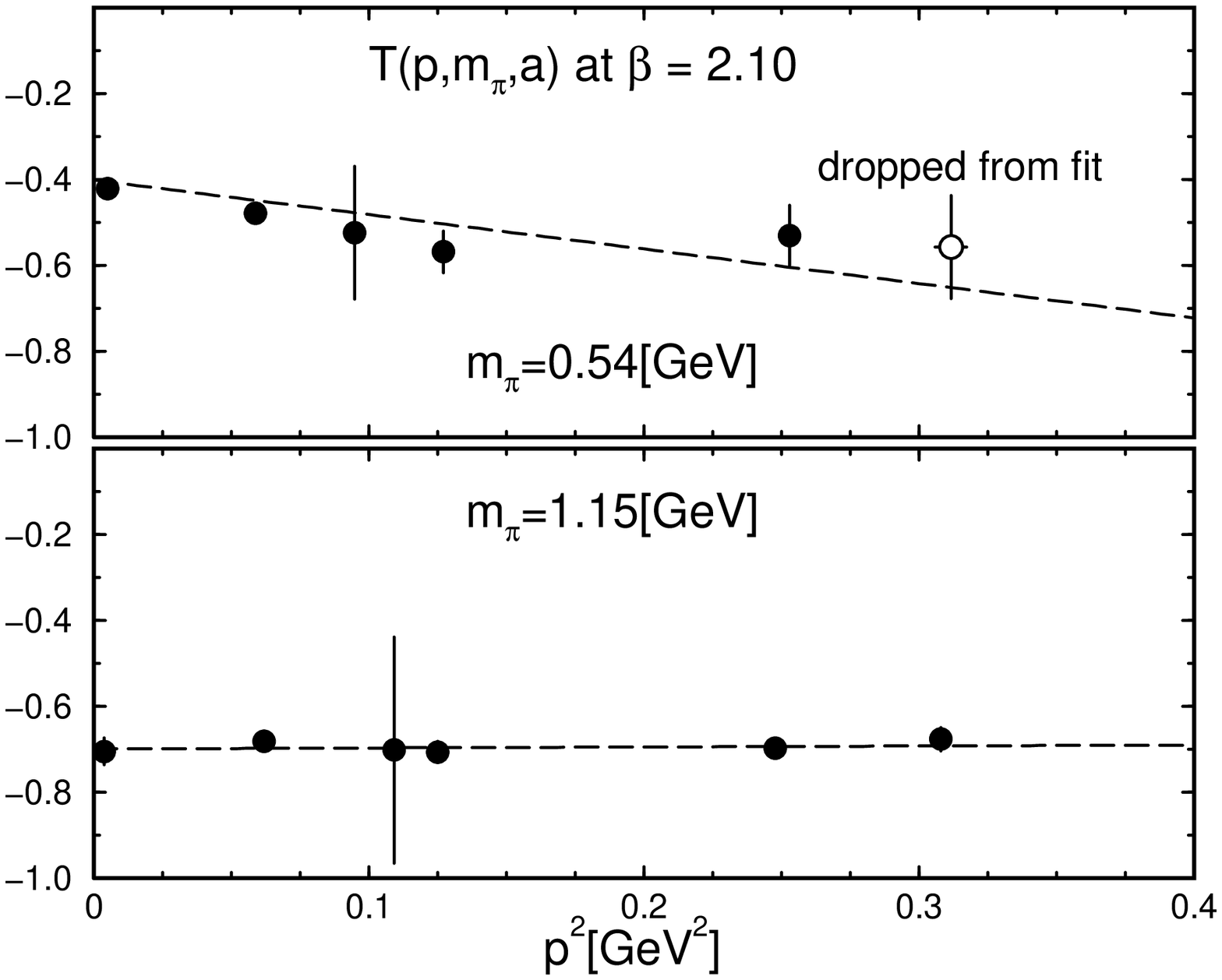}}}
\vspace{-1.cm}
\caption{Scattering amplitudes defined by (1) for the heviest and lightest 
pion masses at $\beta = 1.80$ (top panels) and
$2.10$ (bottom panels).
The dashed lines show the global polynomial fit.
\label{fig:amp}}
\vspace{-6mm}
\end{figure}

Figure \ref{fig:diag} illustrates the effect of diagonalization.
We compare the ratio
$R_n(t) = C_{nn}(t)/G_n^2(t)$, 
where $G_n(t) = \langle 0 | \pi(p_n,t)\pi(-p_n,t_S) | 0 \rangle$
with $p_n^2 = (2\pi/L)^2n$,
before the diagonalization (denoted by open symbols)
to 
$W_n(t) = \lambda_n(t) \cdot G_n^2(t_0) / G_n^2(t)$ 
after the
diagonalization (closed symbols) normalized at $t=t_S$.
For the ground state,  
$W_0(t)\simeq R_0(t)$ so that the  
diagonalization is unnecessary. 
For the $n=1$ state, however, the ratio shows an exponential fall-off
only after diagonalization: $R_1(t)$ blows up at large $t$.
We apply an exponential fit to $W_1(t)$.
The energy eigenvalues for the  $n=0$ and 1 states
change very little even if we adopt the cut-off $N=2$
instead of 3 and 4.

\section{Phase shifts}

We define the `scattering amplitude' 
\begin{equation}
~~~~~~~~~~T(p,m_{\pi},a) = \tan(\delta(p)) E_{\pi} / p
\end{equation}
where $E_{\pi} = \sqrt{ m_{\pi}^2 + p^2 }$.  
This gives the scattering length $a_0$
in the zero-momentum limit:  
$\lim_{p \rightarrow 0} T(p,m_{\pi},a) = a_0 m_{\pi}$.

Figure~\ref{fig:amp} shows the scattering amplitude at
$ \beta=1.80$ and $2.10$ 
for the lightest and heaviest pion masses.
The CM$_n$ refers to the amplitude obtained from the $n$-th state in 
the CM frame, and the L1$_n$ and L2$_n$ to those in the lab frames. 
Note that we have three points between the two CM data,
showing the improvement of the momentum sampling.

We carry out a global fit to the $T$ at each $\beta$ 
with a polynomial 
$T(p,m_{\pi},a) = A_{10}m_{\pi}^2 + A_{20}m_{\pi}^4 + A_{30}m_{\pi}^6
+ A_{01}p^2 + A_{11} m_{\pi}^2 p^2 + A_{21} m_{\pi}^4 p^2 $.
One data point (shown with open symbol) is excluded from the fit, since
the $W_n(t)$ does not show a good exponential behavior.
We find a reasonable fit to the data for both CM 
and lab frames. 

\begin{figure}[t!]
\centerline{\scalebox{.41}[.41]{\includegraphics{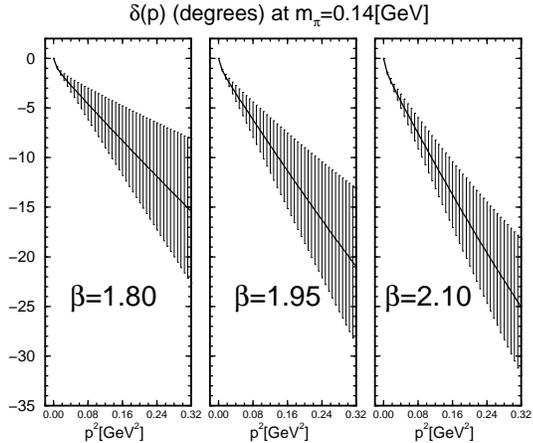}}}
\vspace{-1.0cm}
\caption{Phase shifts at the physical pion mass for the
three values of $\beta$.
\label{fig:phbt}}
\vspace{-8mm}
\end{figure}

In Fig.\ref{fig:phbt} the phase shift at the physical pion 
mass $m_{\pi} = 0.14$ GeV, as calculated from the fit, 
decreases with $\beta$, showing the presence
of an $O(a)$ effect. 

\begin{figure}[t!]
\centerline{\scalebox{.47}[.47]{\includegraphics{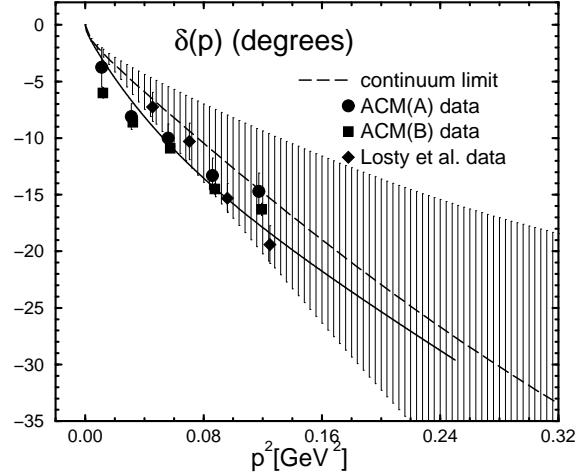}}}
\vspace{-1.2cm}
\caption{Phase shifts in the continuum limit (indicated by 
the dashed curve). The band of bars shows our 
estimate of errors. The solid curve is parametrized by
experimental inputs~\cite{cola} and the symbols are
the experimental data.
\label{fig:phct}}
\vspace{-2mm}
\end{figure}

The continuum limit is taken by extrapolating the 
fits linear in $a$ to the scattering amplitude at the physical pion
mass for each momentum, $T(p,m_{\pi},a) = T(p,m_{\pi}) + a T^{a}(p,m_{\pi})$.
The phase shift in the continuum limit is presented
in Fig.\ref{fig:phct} with the dashed line, associated by  
a band of error bars. This is compared to
the experimental data.

The scattering length obtained by taking the zero-momentum limit of 
$T(p,m_{\pi})$ in the continuum, $ a_0 m_{\pi} = -0.0488(49) $,
may be compared with the prediction of chiral perturbation theory: 
$ a_0 m_{\pi} = -0.0444(10) $~\cite{cola}.

\vspace{2mm}
This work is supported in part by Grants-in-Aid of the Ministry of Education
(Nos.
12304011, 
12640253, 
13135204, 
13640259, 
13640260, 
14046202, 
14740173, 
15204015, 
15540251, 
15540279, 
15740134  
).
The numerical calculations have been carried out 
on the parallel computer CP-PACS.

%

%
%

\end{document}